\documentclass[preprint]{revtex4-1}
\usepackage{epsfig}
\usepackage{graphics}
\usepackage{latexsym}
\usepackage{amsmath}
\usepackage{amssymb}
\usepackage{rotating}
\usepackage{subfigure}
\usepackage{bm}
\usepackage{color}
\usepackage{blindtext}
\usepackage{hyperref}
\usepackage{graphicx}
\begin{document}

\title{Probing nuclear modifications of parton distribution functions through the
isolated prompt photon production at the LHC}

\author{Muhammad Goharipour$^{a}$}
\email{muhammad.goharipour@ipm.ir}

\author{S.~Rostami$^{b}$}
\email{asalrostami.phy@gmail.com}

\affiliation{
$^{a}$School of Particles and Accelerators, Institute for Research in Fundamental Sciences (IPM), PO Box 19568-36681, Tehran, Iran\\
$^{b}$Young Researchers and Elites Club, South Tehran Branch, Islamic Azad University, Tehran, Iran}

\date{\today}

\begin{abstract}

An accurate knowledge of nuclear parton distribution functions (nPDFs) is an essential ingredient of high energy physics calculations when the processes are involving nuclei in the initial state.
It is well known now that the prompt photon production both in hadronic and nuclear collisions is a powerful tool for exploring the parton densities in the nucleon and nuclei especially of the gluon. 
In this work, we are going to perform a comprehensive study of the isolated prompt photon 
production in $ p $-Pb collisions at backward rapidities to find the best kinematic regions in which the experimental measurements have most sensitivity to the nuclear modifications of parton densities. Most emphasis will be placed on the antishadowing nuclear modification. To this aim, we calculate and compare various quantities at different values of center-of-mass energy covered by the LHC and also different rapidity regions to realize which one is most useful.

\end{abstract}


%
\maketitle

\section{Introduction}\label{sec:one} 
Besides the parton distribution functions (PDFs), whether 
unpolarized~\cite{Accardi:2016qay,Ball:2014uwa,Harland-Lang:2014zoa,Jimenez-Delgado:2014twa,Ball:2014uwa,Dulat:2015mca,
Aleedaneshvar:2016dzb,Alekhin:2017kpj,Ball:2017nwa,MoosaviNejad:2016ebo,Khanpour:2016uxh,Butterworth:2015oua} or polarized~\cite{
Jimenez-Delgado:2014xza,Sato:2016tuz,Shahri:2016uzl,Khanpour:2017cha,Ethier:2017zbq,Khanpour:2017fey,Salajegheh:2018hfs}, and fragmentation functions (FFs)~\cite{Nejad:2015fdh,Leader:2015hna,deFlorian:2017lwf,Soleymaninia:2017xhc,Soleymaninia:2018uiv},
the nuclear modifications of PDFs~\cite{Hirai:2007sx,Schienbein:2009kk,Eskola:2009uj,deFlorian:2011fp,Kovarik:2015cma,Khanpour:2016pph,Eskola:2016oht,Wang:2016mzo} 
are also important ingredients of high energy physics calculations,
in particular, for processes involving nuclei in the initial state.
In fact, without having nuclear PDFs (nPDFs) which describe the structure of the colliding nuclei,
the theoretical calculation of the cross sections in any nuclear collision will not be possible.
Thanks to the collinear factorization theorem~\cite{Collins:1989gx,Brock:1993sz}, the nPDFs
can be extracted just in a way similar to the PDFs determination through a global analysis
of nuclear experimental data. 
Nowadays, due to many developments achieved in the phenomenological approaches, 
theoretical calculations and experimental measurements, 
the PDFs are well determined in a wide range of the momentum fraction $ x $.
However, the situation is not very satisfying for the case of nPDFs
because of the lack of experimental data.

Although the main experimental data for constraining nPDFs come from the old fixed-target deep inelastic scattering (DIS) and proton-nucleus Drell-Yan (DY) dilepton
production experiments, there are some analyses in which the neutrino DIS data have also been 
used~\cite{Schienbein:2009kk,deFlorian:2011fp,Eskola:2016oht}. Furthermore, the inclusive pion production from d-Au collisions at RHIC that can be considered as another source to put further constraints on the nuclear gluon distribution is usually used in the nPDFs analyses~\cite{Eskola:2009uj,deFlorian:2011fp,Kovarik:2015cma,Eskola:2016oht}. Recently, EPPS16~\cite{Eskola:2016oht}
has also included, for the first time, the fixed-target DY data in pion-nucleus collisions and 
new LHC proton-lead ($ p $-Pb) data on dijet and heavy gauge-boson production. There are 
also some studies show that important information about nPDFs can be achieved by analyzing the
prompt photon production in nuclear collisions~\cite{Arleo:2007js,BrennerMariotto:2008st,Zhou:2010zzm,Arleo:2011gc,Helenius:2014qla,Goharipour:2017uic}, jet and dijet photoproduction measurements at a future electron-ion collider (EIC)~\cite{Klasen:2017kwb,Klasen:2018gtb}, single inclusive jet production at very forward rapidity~\cite{Bury:2017xwd}, and heavy-flavor production in $ p $-Pb collisions~\cite{Kusina:2017gkz}. 

The main reason for such efforts to include more accurate experimental data 
as much as possible in the global analysis of nPDFs 
is to achieve more valid nuclear modifications of PDFs with less uncertainties. In fact, most of experimental data that are used in the nPDFs analyses can
just put well constraints on the quark nuclear modifications at fairly large values of $ x $, while the sea quarks and gluon distributions cannot be controlled as well, especially at 
smaller values of $ x $. Actually, the limited kinematic reach of data leads to a crucial difference
between the results obtained by various groups for the gluon nuclear modifications and also
their uncertainties. Therefore, since the common constraint of nuclear gluon distribution that comes from the DGLAP evolution at higher order of perturbation theory is not enough to make accurate theoretical predictions of the physical observables, the inclusion of the experimental data that are directly sensitive to the gluon density in the nPDFs analysis is inevitable.

It is well known now that the direct photon production in hadronic collisions
is one of the excellent tools for studding the large momentum transfer processes~\cite{Aurenche:1983ws,Owens:1986mp,Huston:1995vb,Aurenche:1998gv,Aurenche:2006vj} and then testing the perturbative Quantum Chromodynamics (pQCD), since photons couple in a point-like fashion to the quark constituents of the colliding hadrons. The direct photon production is also recognized as a useful tool~\cite{Alam:1999sc} for studding the whole time evolution and dynamics of the deconfined, strongly interacting matter namely the quark gluon plasma (QGP)~\cite{Shuryak:1980tp,dEnterria:2006mtd} created in heavy-ion collisions. It is worth noting in this context that in
nucleus-nucleus collisions, direct photon production contains two kinds of photons: thermal photons and prompt photons~\cite{Paquet:2016dry}. The first ones have a thermal origin and the second ones come from cold processes.
By definition, the prompt photons do not include the photons coming
from the decays of hadrons such as $ \pi_0 $, $ \eta $ produced in the collision.
The measurement of the prompt photon production in hadronic collisions can bring very useful information of gluon PDFs~\cite{Aurenche:1988vi,Vogelsang:1995bg,Ichou:2010wc,dEnterria:2012kvo,Campbell:2018wfu} and, as mentioned before, the gluon nuclear modifications~\cite{Arleo:2007js,BrennerMariotto:2008st,Zhou:2010zzm,Arleo:2011gc,Helenius:2014qla,Goharipour:2017uic} in the case of nuclear collisions.
Such measurements have another advantages in various areas of the high energy physics, for example
for searching the intrinsic heavy quark components of the nucleon if photons are associated with a heavy quark~\cite{Bednyakov:2013zta,Rostami:2016dqi}.
Another important point should be mentioned is that in order to reject the background of photons
that are not considered as prompt photons, an isolation criterion is usually used. 

In view of the experimental efforts, there are various measurements of direct photon production in heavy-ion collisions by different collaborations. For example, on can refer to the PHENIX Collaboration measurements in Au-Au~\cite{Adler:2005ig,Afanasiev:2012dg,Adare:2008ab,Adare:2014fwh} and d-Au~\cite{Adare:2012vn} collisions at RHIC and center-of-mass energy of $ \sqrt{s}=200 $ GeV, and also the ALICE Collaboration measurements~\cite{Wilde:2012wc,Adam:2015lda} at the LHC from Pb-Pb collisions at $ \sqrt{s}=2.76 $ TeV. For the case of isolated prompt photon production in Pb-Pb collisions, there are also some measurements by the ATLAS~\cite{Aad:2015lcb} and CMS~\cite{Chatrchyan:2012vq} Collaborations at $ \sqrt{s}=2.76 $ TeV. Despite all these experimental efforts, another kind of measurements that is very important for achieving accurate information of the gluon nuclear modifications is the measurement of the prompt photon production in $ p $-A collisions especially at higher values of the center-of-mass energy and forward rapidities~\cite{Goharipour:2017uic}. Recently, the ALICE Collaboration has reported the measurement of inclusive production cross section for isolated prompt photons in $ p $-Pb collisions at $ \sqrt{s}=8.16 $ TeV as a function of photon
transverse energy from 25 to 500 GeV and over nearly five units of 
pseudorapidity~\cite{ATLAS:2017ojy}.

In this work, we are going to perform a comprehensive study of the isolated prompt photon 
production in $ p $-Pb collisions at different values of center-of-mass energy covered by the LHC and also different rapidity regions. Since the study of the shadowing nuclear modification of parton densities through the prompt photon production at the LHC has been previously done in some papers, in the present study more emphasis will be placed on the antishadowing nuclear modification. In fact, the main goal of this work is finding the best kinematic regions in which the measurement of the isolated prompt photon production in $ p $-Pb collisions has most sensitivity to the antishadowing area and then is more useful to constrain the antishadowing nuclear modification of the gluon distribution.

The contents of the present paper are as follows. In Sec.~\ref{sec:two}, we
discuss the nuclear modifications of PDFs with emphasis on gluon nuclear modification for Pb-nucleus
and compare predictions of various phenomenological groups at factorization scales corresponding to energies covered by the LHC. Our framework and anything needed to calculate the isolated prompt photon production cross sections and other related quantities is briefly described in Sec.~\ref{sec:three}. 
Sec.~\ref{sec:four} is devoted to study the isolated prompt photon production in $ p $-Pb collision
at different values of center-of-mass energy and different rapidity regions to find the best kinematic regions for constraining the antishadowing nuclear modification of gluon PDF. In this regard, various quantities, including the nuclear modification ratio, the rapidity and energy normalized nuclear modification ratios, and the ratio of nuclear cross sections for different rapidity regions and also different energies, are calculated and compared with each other to realize which one is most useful. We also present a comparison between the predictions obtained from various nPDFs and study the uncertainties due to scale variations and nPDFs. Finally, we summarize our results and conclusions in Sec.~\ref{sec:five}.

%
\section{The nuclear modifications of parton densities}\label{sec:two}
As mentioned in the Introduction, in order to calculate any cross section of the particles production in high energy nuclear collisions, it is necessary to have accurate nPDFs which describe 
the structure of the colliding nuclei. The determination of the nuclear modifications of
parton densities usually has a similar procedure to the case of PDFs of free nucleon
that is performing a QCD global analysis of nuclear experimental data~\cite{Hirai:2007sx,Schienbein:2009kk,Eskola:2009uj,deFlorian:2011fp,Kovarik:2015cma,Khanpour:2016pph,Eskola:2016oht,Wang:2016mzo}.
Regardless of this fact that the determination of nPDFs with a precision similar to the PDFs 
is not possible at the present due to the lake of nuclear experimental data, the different 
phenomenological approaches used by various groups to extract the nuclear modifications of
parton densities have led to some considerable differences in the results obtained,
both in behavior and uncertainty, especially for the case of gluon and sea quarks nuclear modifications. The differences between the nPDFs from various phenomenological groups, in turn, can lead to the different results for the theoretical predictions of the physical observables. 
Therefore, in inverse, the accurate measurements of some particle production cross sections in
nuclear collisions can bring us new useful information to judge about the behavior of nuclear modifications in different kinematic regions and also decrease their uncertainties. The measurements
of the prompt photon production is one of the most useful ones to obtain information about the
gluon nuclear modifications~\cite{Arleo:2007js,BrennerMariotto:2008st,Zhou:2010zzm,Arleo:2011gc,Helenius:2014qla,Goharipour:2017uic}, since the gluon directly contribute to it 
through the Compton scattering $ q(\bar q)g \rightarrow \gamma q(\bar q) $ subprocesses at the Born-level.

From past to present, several QCD global analyses of nPDFs have been done which are different from
various aspects such as the amount and kind of the experimental data included, the perturbative order 
of the analysis, the parametrization of the nuclear modifications, the method for calculating uncertainties and treatment of the heavy quark masses (for review see Refs.~\cite{Goharipour:2017uic,Paakkinen:2018zbs}). In most of the analyses performed so far, like the HKN07~\cite{Hirai:2007sx}, EPS09~\cite{Eskola:2009uj}, DSSZ~\cite{deFlorian:2011fp}, 
KA15~\cite{Khanpour:2016pph} and EPPS16~\cite{Eskola:2016oht} analyses, the bound-proton PDFs have been defined in terms of nuclear modification factors $ R_i^A $ that characterize the relation of the  PDFs of a proton inside a nucleus with atomic number $ A $, $ f_i^{p/A} $, with respect to free-proton PDFs $ f_i^p $,
\begin{equation}
R_i^A(x,Q^2)\equiv\frac{ f_i^{p/A}(x,Q^2)}{ f_i^p(x,Q^2)}.
\label{eq1}
\end{equation}
In this approach, for determining the nuclear modifications one needs to fix the free-proton PDFs form an analysis already performed on the nucleon experimental data~\cite{Accardi:2016qay,Ball:2014uwa,Harland-Lang:2014zoa,Jimenez-Delgado:2014twa,Dulat:2015mca,
Aleedaneshvar:2016dzb,Alekhin:2017kpj,Ball:2017nwa,MoosaviNejad:2016ebo,Khanpour:2016uxh,Butterworth:2015oua}. However, there is another approach in which the nPDFs are parametrised directly as a function of $ x $ at the starting scale $ Q_0^2 $ without any factorisation into a nuclear modification factor and free-proton PDFs. In this approach that has been used by the nCTEQ group~\cite{Schienbein:2009kk,Kovarik:2015cma}, the $ A $-dependence of nPDFs is
introduced in the coefficients of their functional form.

\begin{figure}[h!]
\includegraphics[width=0.53\textwidth]{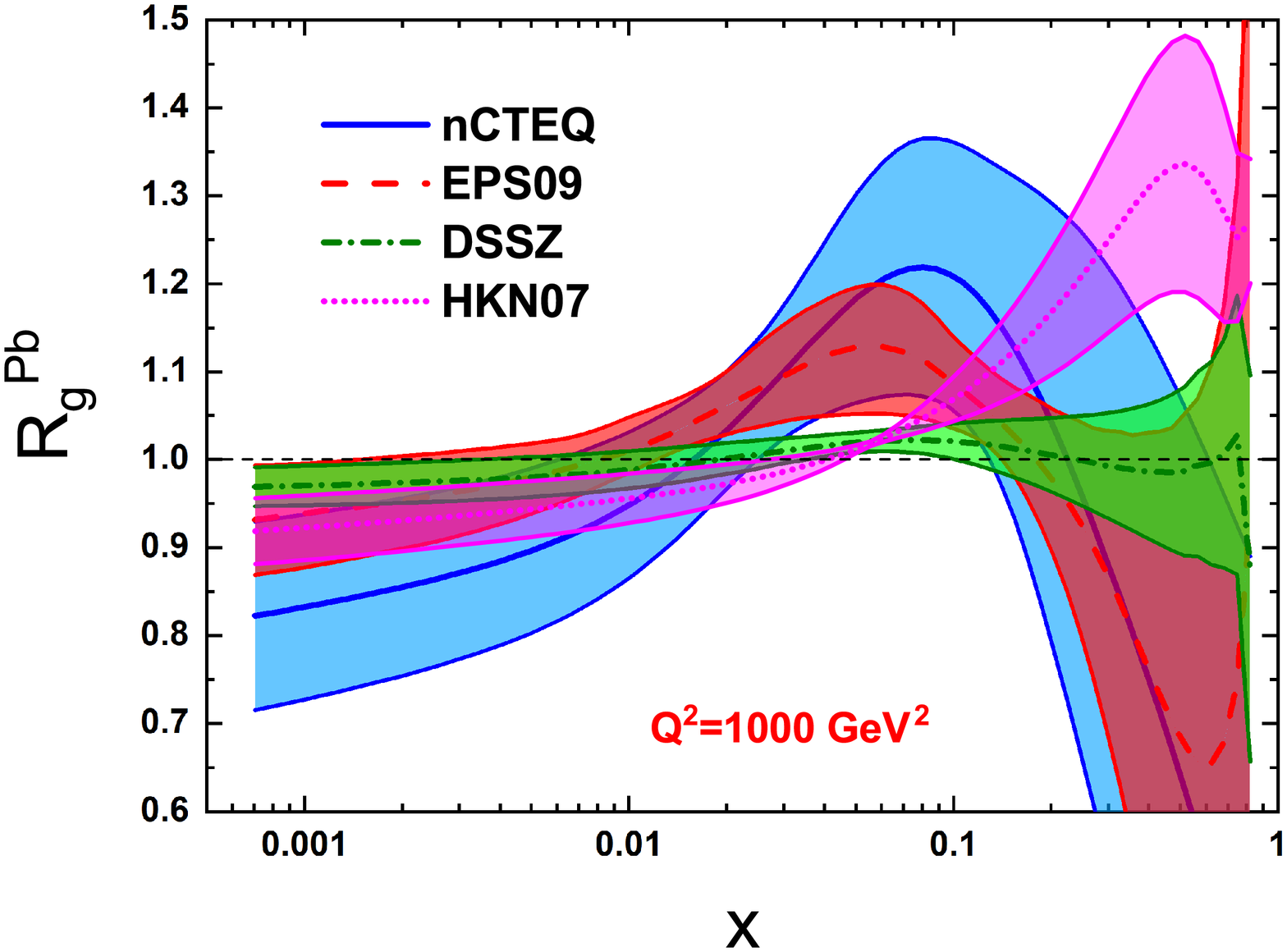}
\includegraphics[width=0.56\textwidth]{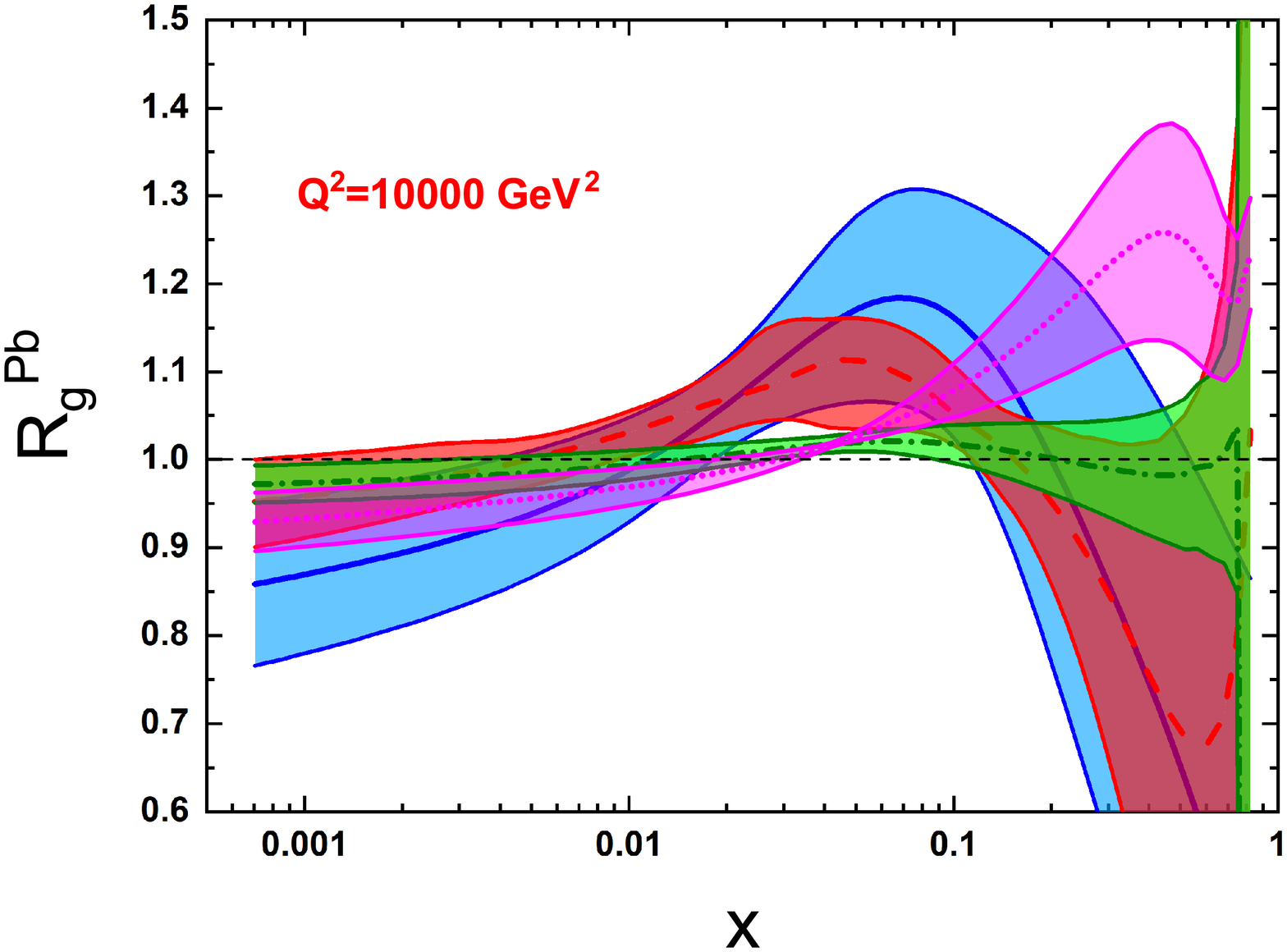}
\caption{ A comparison
between the nuclear modifications of the gluon PDF in a Pb-nucleus with their uncertainties from 
the nCTEQ15~\cite{Kovarik:2015cma} (blue solid), EPS09~\cite{Eskola:2009uj} (red dashed), 
DSSZ~\cite{deFlorian:2011fp} (green dashed-dotted) and HKN07~\cite{Hirai:2007sx} (pink dotted)
at $ Q^2= $1000 (top panel) and 10000 (bottom panel) GeV$ ^2 $.  }
\label{fig:fig1}
\end{figure}

A comparison between nuclear modifications of the gluon PDF in a Pb-nucleus with their uncertainties from the nCTEQ15~\cite{Kovarik:2015cma} (blue solid), EPS09~\cite{Eskola:2009uj} (red dashed), 
DSSZ~\cite{deFlorian:2011fp} (green dashed-dotted) and HKN07~\cite{Hirai:2007sx} (pink dotted) has been shown in Fig.~\ref{fig:fig1}. The comparison has been made for two different scales 
$ Q^2= $1000 (top panel) and 10000 (bottom panel) GeV$ ^2 $ which are covered by the LHC
energies. As can clearly seen, almost in whole values of $ x $, there are considerable differences between the results of various groups in both best fits and uncertainties. Overall, the
nuclear modification of parton densities, in terms of $ x $, is usually divided to four areas,
namely, the shadowing, antishadowing, EMC effect and Fermi motion~\cite{Paakkinen:2018zbs}. The antishadowing area, that is the favorite case of the present study, is attributed to an enhancement
in $ R_i^A $ around $ x\sim0.1 $. The decreasing areas before and after it are named as shadowing and 
EMC effect. There is another increasing area at very large values of $ x $ which is referred to Fermi motion. For the case of nCTEQ15, we see some stronger shadowing, antishadowing and EMC effect
in gluon nuclear modification compared to other groups that makes it a favorite case 
to use for studding in this subject. Note also that the nCTEQ15 prediction has wider uncertainty than other groups in all values of $ x $. Focusing on the antishadowing area, one can see that
the EPS09 has a similar treatment to nCTEQ15 but somewhat softer. The DSSZ dose not show any remarkable antishadowing nuclear modification in gluon density and for the case of HKN07, it appears at some larger values of $ x $ in analogy to the nCTEQ15 and EPS09. Accordingly,
one can easily conclude that such differences can lead to different results for predictions of any process at nuclear collisions which is sensitive to antishadowing nuclear modification of the gluon density.

In the prompt photon production at the LHC, RHIC or any hadron collider, various $ x $ regions can be explored in the PDFs of target and projectile according to following approximate relation
\begin{equation}
x_{1,2}\approx \frac{2p_\textrm{T}}{\sqrt s} e^{\pm \eta},
\label{eq2}
\end{equation}
which clearly depends on the center-of-mass energy $ \sqrt s $ of collision, 
and also the transverse momentum $ p_\textrm{T} $ and pseudorapidity $ \eta $ of photons.
According to Eq.~\ref{eq2}, for the case of $ p $-Pb collisions at LHC with high values of $ \sqrt s $, we can only investigate the antishadowing nuclear modification at backward rapidities.
Actually, at forward rapidities, the smaller values of $ x_2 $ can be probed which are
related to the shadowing area as studied before~\cite{Arleo:2007js,BrennerMariotto:2008st,Zhou:2010zzm,Arleo:2011gc,Helenius:2014qla,Goharipour:2017uic}. In this work, we are looking
for the best kinematic regions in which the isolated prompt photon production at
the LHC $ p $-Pb collisions has most sensitivity to the antishadowing nuclear modification and then more potential to constrain it. 

%
\section{The framework of the calculations}\label{sec:three}
As seen in the previous section, the gluon nuclear modifications extracted by various phenomenological groups have major differences almost in all regions of the momentum fraction $ x $. This fact can lead to the different results for any physical observable which is dependent on the nPDFs.
As mentioned before, in this work we are going to use the isolated prompt photon production in $ p $-Pb collisions at the LHC as a useful tool for investigating the gluon nuclear modifications, in order to find the best kinematic regions that have the most sensitivity to the antishadowing area. In this section, we briefly discuss the framework of the calculation of the cross sections. 

The prompt photon production has been an interesting subject of research more than three decades~\cite{Arleo:2007js,BrennerMariotto:2008st,Zhou:2010zzm,Arleo:2011gc,Helenius:2014qla,Goharipour:2017uic,Aurenche:1983ws,Aurenche:1987fs,Owens:1986mp,Huston:1995vb,Aurenche:1989gv,Baer:1990ra,Berger:1990et,Gordon:1994ut,Aurenche:1998gv,Catani:2002ny,Aurenche:2006vj,Belghobsi:2009hx,Schwartz:2016olw,Lipatov:2016wgr,Helenius:2013bya,Fontannaz:2001ek,Jezo:2016ypn,Campbell:2016lzl,Siegert:2016bre,Benic:2016uku,Kohara:2015nda,Klasen:2017dsy,Goharipour:2017rjl,Aurenche:1988vi,Vogelsang:1995bg,Ichou:2010wc,dEnterria:2012kvo,Campbell:2018wfu}. The prompt photons (those arise from processes during the collision not those produced from the decay of hadrons) consist of two types of photons: 
direct and fragmentation photons. The conceptual difference of the direct and fragmentation photons is related to their production procedure. Actually, direct photons are produced directly from
initial hard scattering processes of the colliding quarks or gluons, while fragmentation photons 
are produced from the fragmentation of high-$ p_\textrm{T} $ quarks and gluons which themselves have already produced during the hard collisions (for more information see, for example, Refs.~\cite{Catani:2002ny,Aurenche:2006vj}). The theoretical calculations of the direct and fragmentation components can be performed completely separate so that the prompt photon production cross
section in hadronic collisions can be written as follows:
\begin{equation}
d\sigma_{h_1h_2}^{\gamma +X}= d\sigma_{h_1h_2}^{\textrm{D~} \gamma +X}+d\sigma_{h_1h_2}^{\textrm{F~} \gamma +X},
\label{eq3}
\end{equation}
where D and F refer to the direct and fragmentation parts, respectively.
Note that the notation $ X $ is introduced to indicate the inclusive nature of the cross section.
The cross section of the prompt photon production at hadronic collisions can be calculated through
some computer codes such as J\textsc{et}P\textsc{hox}~\cite{Catani:2002ny,Aurenche:2006vj,Belghobsi:2009hx} and P\textsc{e}T\textsc{e}R~\cite{Becher:2013vva} up to NLO. The next-to-next-to leading order (NNLO) calculations of the direct photon production have also been presented recently~\cite{Campbell:2016lzl}. In the present study, we use J\textsc{et}P\textsc{hox} for performing all calculations which is a Monte Carlo program written as a partonic event
generator for the prediction of processes with photons in
the final state, and has also various facilities, for example using different procedures for isolating photons as we discuss in the following. 

One of the important points should be considered in the calculation of the cross section of Eq.~\eqref{eq3} is the values of the renormalization $ \mu $, (initial state) factorization $ M $, and fragmentation $ M_F $ scales. The renormalization and factorization scales appear both in the direct and fragmentation components of the cross section, but the fragmentation scale appears only 
in the fragmentation part. Overall, due to scale variations, one can consider some uncertainties
in theoretically calculated results (for more information about the scale uncertainties
in the prompt photon production in hadronic collisions see, for example, Ref.~\cite{Goharipour:2017rjl}). In all calculations which are presented in the next section, the renormalization, factorization and fragmentation scales are set to the photon transverse momentum $ \mu=M=M_F=p_\textrm{T}^\gamma $, except the last subsection in which we study the uncertainties due to scale variations. To calculate the cross section of Eq.~\eqref{eq3} one also needs to use the free-proton PDFs and FFs of the photons. Actually, the PDFs are used for calculating both the direct and fragmentation components, but the FFs are only used in the calculation of the fragmentation component. On the other hand, if we are interested to the proton-nucleus or nucleus-nucleus collisions, we need also the nuclear modification of PDFs for each nucleus contributing in the collision. In this work, we use
the set II of the NLO Bourhis-Fontannaz-Guillet (BFG) FFs of photons~\cite{Bourhis:1997yu} and
the NLO PDF stes of CT14~\cite{Dulat:2015mca} with $ \alpha_s(M_Z)=0.118 $ by virtue of the LHAPDF package~\cite{Buckley:2014ana}. For the nuclear modification of PDFs, we choose the nCTEQ15 nPDFs~\cite{Kovarik:2015cma} that show stronger shadowing, antishadowing and EMC effect nuclear modifications according to Fig.~\ref{fig:fig1}. However, in the end of next section, we also present a comparison between the nCTEQ15, EPS09~\cite{Eskola:2009uj}, DSSZ~\cite{deFlorian:2011fp} and HKN07~\cite{Hirai:2007sx} predictions.

At hadronic collisions, different subprocesses are contributing to the prompt photon production.
In the case of LO approximation, they are the quark-gluon Compton scattering $ q(\bar q)g \rightarrow \gamma q(\bar q) $ and quark-antiquark annihilation $ q\bar q \rightarrow \gamma g $ which both are
Born-level subprocesses. At NLO, the situation is some more complicated, since in addition to the $ q(\bar q)g \rightarrow \gamma gq(\bar q) $ and $ q\bar q \rightarrow \gamma gg $, there are some subprocesses from the virtual corrections to the Born-level processes. However, because of the point-like coupling of the photon to quarks, such a calculation is more easier than other 
processes~\cite{Aurenche:1983ws,Aurenche:1987fs,Fontannaz:2001ek}. It is worth pointing out that
at $ pp $ colliders such as LHC and RHIC, the $ q\bar q $ annihilation channel is suppressed compared to the other subprocesses. This is the main reason for this fact that the prompt photon production
at the LHC provides direct information on the gluon distribution, especially at smaller values of $ x $, because it is dominant rather than sea quarks at this region. The $ q\bar q $ channel
becomes more important at the Tevatron that is a $ p\bar p $ collider~\cite{Ichou:2010wc}.
Another points should be mentioned, according to Eq~\ref{eq2}, is that the photon production at the
LHC probes values of $ x $ that are considerably smaller than at the Tevatron. In the present 
study, we include all diagrams up to LO and NLO order of QED and QCD coupling, respectively,
defined in the $ \overline{\textrm{MS}} $ renormalization scheme for calculating the prompt photon production cross section. Note also that the fine-structure constant ($ \alpha _{\textrm{EM}}$) 
is set to 1/137.

In the experimental view, if we are interested to measure the prompt photon production inclusively,
the background of secondary photons coming from the decays of hadrons such as $ \pi_0 $, $ \eta $
should be well rejected, since they are not prompt photons by definition. The best idea for doing
it is imposing an isolation criterion for photons. Although there are different isolation criteria
can be used to isolate photons~\cite{Catani:2002ny,Kunszt:1992np,Glover:1993xc,Frixione:1998jh,Hall:2018jub}, the most used is the cone criterion~\cite{Catani:2002ny} which is implementable also at the partonic level. The cone isolation criterion considers a photon as an isolated photon if, inside a cone of radius $ R $ which is defined as $ (y-y_\gamma)^2+(\phi-\phi_\gamma)^2\leq R^2 $ in terms of rapidity $ y $ and azimuthal angle $ \phi $ around the photon direction, the amount
of hadronic transverse energy $ E_\textrm{T}^{\textrm{had}} $ is smaller than some value
$ E_\textrm{T}^{\textrm{~max}} $:
\begin{equation}
E_\textrm{T}^{\textrm{had}}\leq E_\textrm{T}^{\textrm{~max}}.
\label{eq4}
\end{equation}
The value of $ R $ is usually set to $ R=0.4 $ in the experimental analyses of isolated prompt photon production~\cite{Chatrchyan:2012vq,Chatrchyan:2011ue,Aad:2013zba,Aad:2016xcr,Aaboud:2017cbm,ATLAS:2017ojy}.
However, $ E_\textrm{T}^{\textrm{had}} $ is considered both as a fixed value or a fixed fraction
of the transverse momentum of the photon $ p_\textrm{T}^\gamma $ and a function of $ p_\textrm{T}^\gamma $. It should be also noted that the photon isolation significantly reduces the fragmentation components of the prompt photon cross section, since the isolation cut discards the prompt photon events that have too much hadronic activity around the photon~\cite{Catani:2002ny}.
In all calculations which are presented in this paper, we use a tighter isolation cut, 
$ E_\textrm{T}^{\textrm{had}}< 2 $, with $ R=0.4 $.

Another point should be considered is the theoretical uncertainties in
the results with respect to the various sources including the PDFs, nPDFs, scales and FFs 
uncertainties. For more information about the impact of these sources on the uncertainties
of the prompt photon production cross section at the LHC, one can refer, for example, to 
Ref.~\cite{Goharipour:2017uic}. It can be shown that the dominant source of theoretical uncertainty 
is that arising from the scale uncertainties and the uncertainty due to free-proton PDFs and FFs 
uncertainties can be ignored compared to the nPDFs uncertainties. Although our main goal is performing a comprehensive study of the isolated prompt photon production in $ p $-Pb collisions to find the best kinematic regions for constraining the antishadowing nuclear modifications of PDFs, we study in the next section the nPDFs uncertainties using the nCTEQ15~\cite{Kovarik:2015cma} nPDFs, in addition to scale uncertainties.

It is well know now that the minimum bias nuclear modification ratio is a
very useful quantity for studding the nuclear modifications of parton densities~\cite{Arleo:2011gc,Helenius:2014qla,Goharipour:2017uic}. It is defined as the ratio of the cross section
of collisions containing nucleus to the cross section of pure proton-proton collisions. For the case of prompt photon production in $ p $-Pb collisions we have,
\begin{equation}
R_{p\textrm{Pb}}^\gamma\equiv \frac{d\sigma/dp_\textrm{T} (p\textrm{+Pb}\rightarrow \gamma +\textrm{X})}{208\times d\sigma/dp_\textrm{T} (p+p\rightarrow \gamma +\textrm{X})}.
\label{eq5}
\end{equation}
Such a quantity is clearly more sensitive to the nuclear modifications of PDFs.
Moreover, many sources of theoretical and experimental uncertainties are canceled to a large extent
in the calculation and measurement of nuclear modification ratio $ R_{p\textrm{Pb}}^\gamma $.
In theoretical side, we can refer to the PDFs, FFs and scales uncertainties, since there are both
in numerator and denominator. Actually, the main source of theoretical uncertainty in
$ R_{p\textrm{Pb}}^\gamma $ comes from the nPDFs uncertainties. Another point should be
considered is that $ R_{p\textrm{Pb}}^\gamma $ is not normalized to 1 when
no nuclear modifications in the parton densities are assumed because of the isospin effect.
In the next section, we begin our study by calculating the nuclear modification ratio for isolated prompt photon production in $ p $-Pb collisions at different values of center-of-mass energy. 
Then, we introduce and calculate another possible quantities which can be useful for investigating the nuclear modifications of PDFs.
%
%
\section{Study of isolated prompt photon production at the LHC}\label{sec:four}
After introducing the framework of the calculations, we are now in position to calculate the isolated prompt photon production in $ p $-Pb collisions in order to investigate the nuclear modifications
of parton densities, especially of the gluon. As mentioned before, we are looking for the best kinematic regions in which the measurement of the isolated prompt photon production in $ p $-Pb collisions has most sensitivity to the antishadowing area. 
To this aim, in the following subsections, we calculate and compare various quantities, including the nuclear modification ratio, the rapidity and energy normalized nuclear modification ratios, and the ratio of nuclear cross sections for different rapidity regions and also different energies to realize which one is most useful.

\subsection{Nuclear modification ratio at backward rapidities}

As a first step, we calculate the nuclear modification ratio $ R_{p\textrm{Pb}}^\gamma $ defined
in Eq.~\ref{eq5} for $ p $-Pb collisions at $ \sqrt{s}=2.76 $ TeV. As mentioned before, at such
energies and higher ones, it is not possible to investigate the antishadowing nuclear modification
at forward rapidities since according to Eq.~\ref{eq2}, the cross section is sensitive to
parton densities of target at smaller momentum fraction $ x_2 $.

\begin{figure}[t!]
\centering
\includegraphics[width=0.65\textwidth]{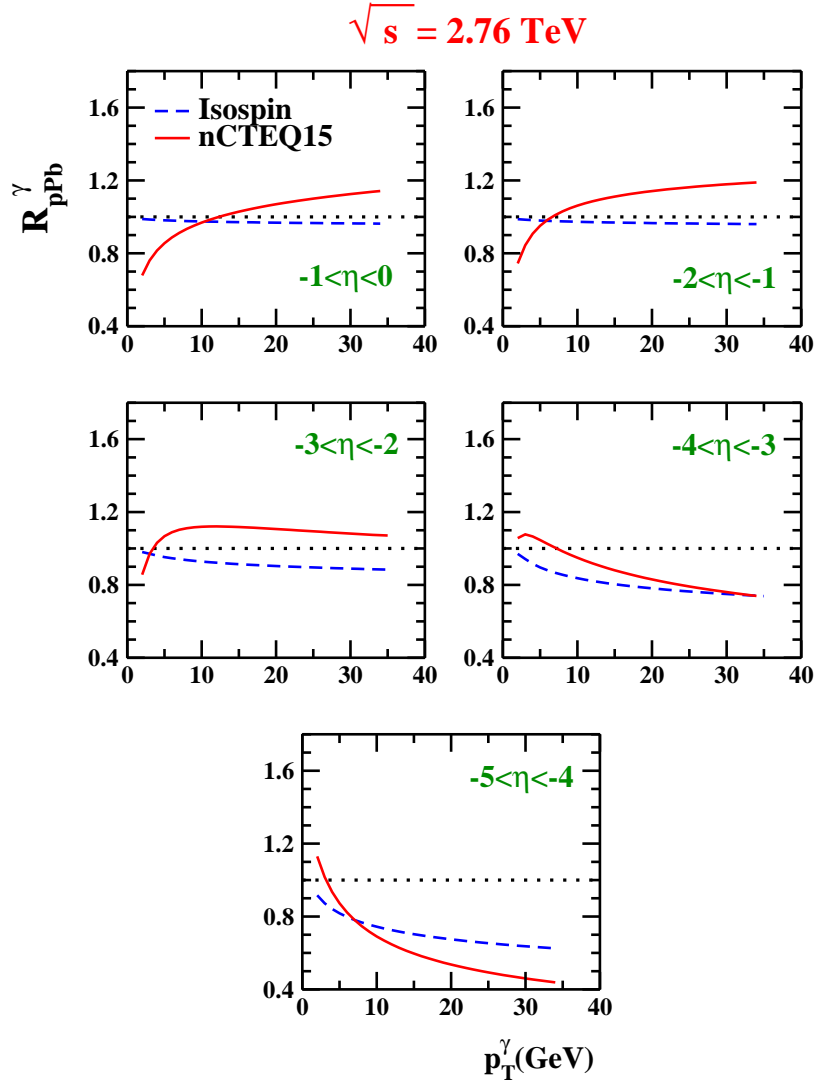}
\caption{The nuclear modification ratio $ R_{p\textrm{Pb}}^\gamma $
as a function of $ p_\textrm{T}^\gamma $ for $ p $-Pb collisions at $ \sqrt{s}=2.76 $ TeV and five different rapidity regions in backward interval $ -5 < \eta^\gamma < 0 $, with no nuclear modifications in the PDFs (blue dashed curve) and using the nCTEQ15~\cite{Kovarik:2015cma} nuclear modifications (red solid curve).}
\label{fig:fig2}
\end{figure}

Fig.~\ref{fig:fig2} shows the results obtained as a function of photon transverse momentum $ p_\textrm{T}^\gamma $ at five different (pseudo)rapidity regions in backward interval $ -5 < \eta^\gamma < 0 $. In each panel, we have shown both the calculation with no nuclear modifications in the PDFs which indicates only the isospin effect (blue dashed curve) and the calculation obtained using the nCTEQ15~\cite{Kovarik:2015cma} nuclear modifications (red solid curve) in order to separate the genuine nuclear effects from the sheer isospin effect.
According to results obtained for $ \sqrt{s}=2.76 $ TeV, as the absolute value of rapidity increases in backward region, the nuclear modification ratio $ R_{p\textrm{Pb}}^\gamma $ moves from the shadowing area to antishadowing and then the EMC effect.
It can be clearly seen that the best kinematic regions for constraining antishadowing nuclear
modification are $ -2 < \eta^\gamma < -1 $ and $ -3 < \eta^\gamma < -2 $, in which
the nCTEQ15 prediction places above the isospin baseline almost in most areas of the 
$ p_\textrm{T}^\gamma $ spectrum. To be more precise, at some values of $ p_\textrm{T}^\gamma $, the difference between the isospin baseline and nCTEQ15 prediction even reaches 20\%. It should be noted that the antishadowing nuclear
modification can also be investigated from $ -1 < \eta^\gamma < -0 $ and $ -3 < \eta^\gamma < -4 $
regions at larger and smaller values of $ p_\textrm{T}^\gamma $, respectively, 
but not so well as the case of $ -2 < \eta^\gamma < -1 $ and $ -3 < \eta^\gamma < -2 $ regions.
 
\begin{figure}[t!]
\centering
\includegraphics[width=0.65\textwidth]{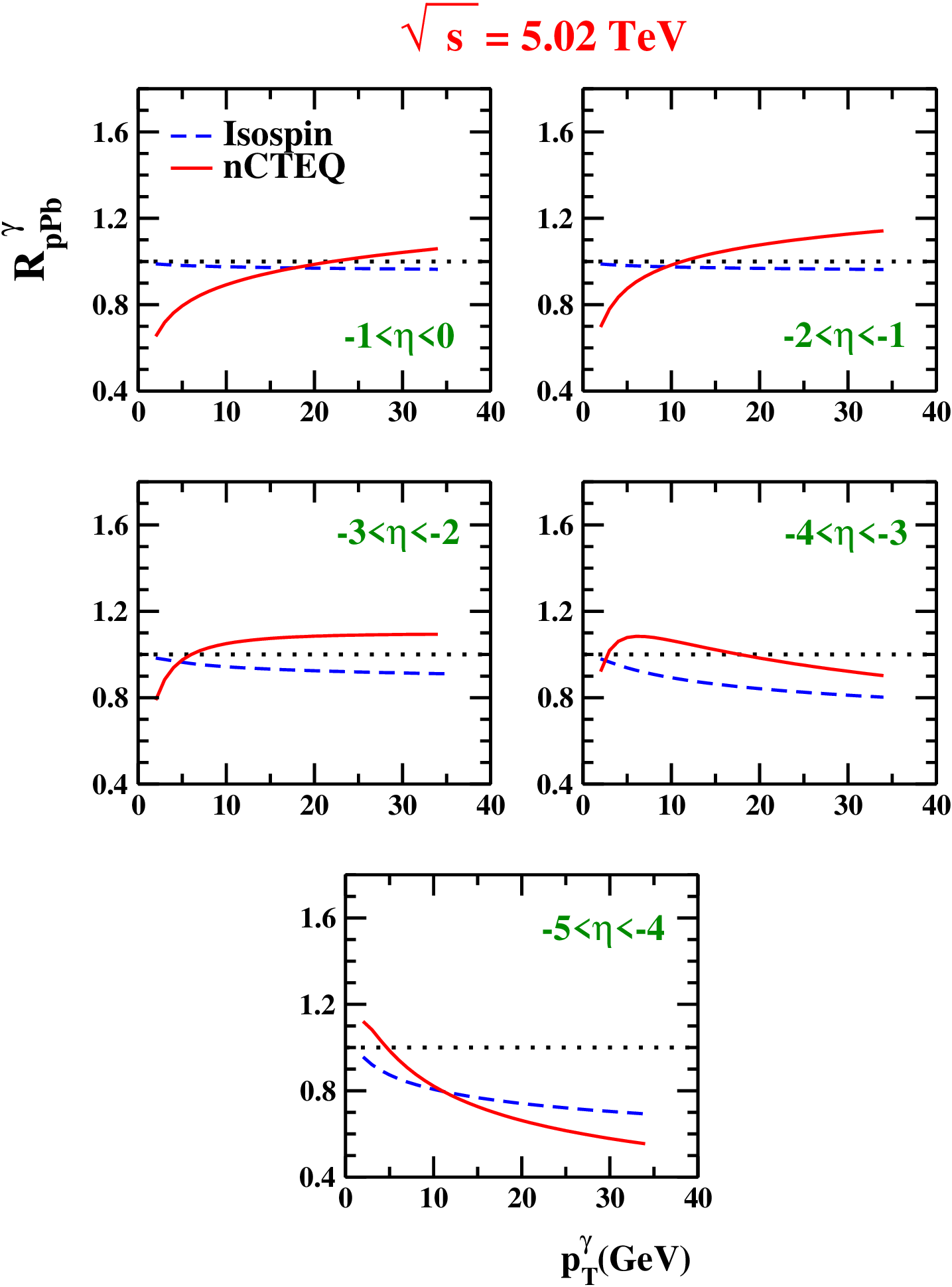}
\caption{Same as Fig.~\ref{fig:fig2}, but for $ \sqrt{s}=5.02 $ TeV.}
\label{fig:fig3}
\end{figure}

Fig.~\ref{fig:fig3} shows the same results as Fig.~\ref{fig:fig2}, but for the center-of-mass energy of $ \sqrt{s}=5.02 $ TeV. As one can see, when the value of $ \sqrt{s} $ increases, the cross section and then the nuclear modification ratio $ R_{p\textrm{Pb}}^\gamma $ tends to smaller values of $ x $.
Therefore, compared to the previous case, moving from the shadowing area towards antishadowing and EMC effect happens somewhat later as the absolute value of rapidity increases. In this case, the best kinematic regions for constraining antishadowing nuclear modification are $ -3 < \eta^\gamma < -2 $ and $ -4 < \eta^\gamma < -3 $. 

\begin{figure}[t!]
\centering
\includegraphics[width=0.65\textwidth]{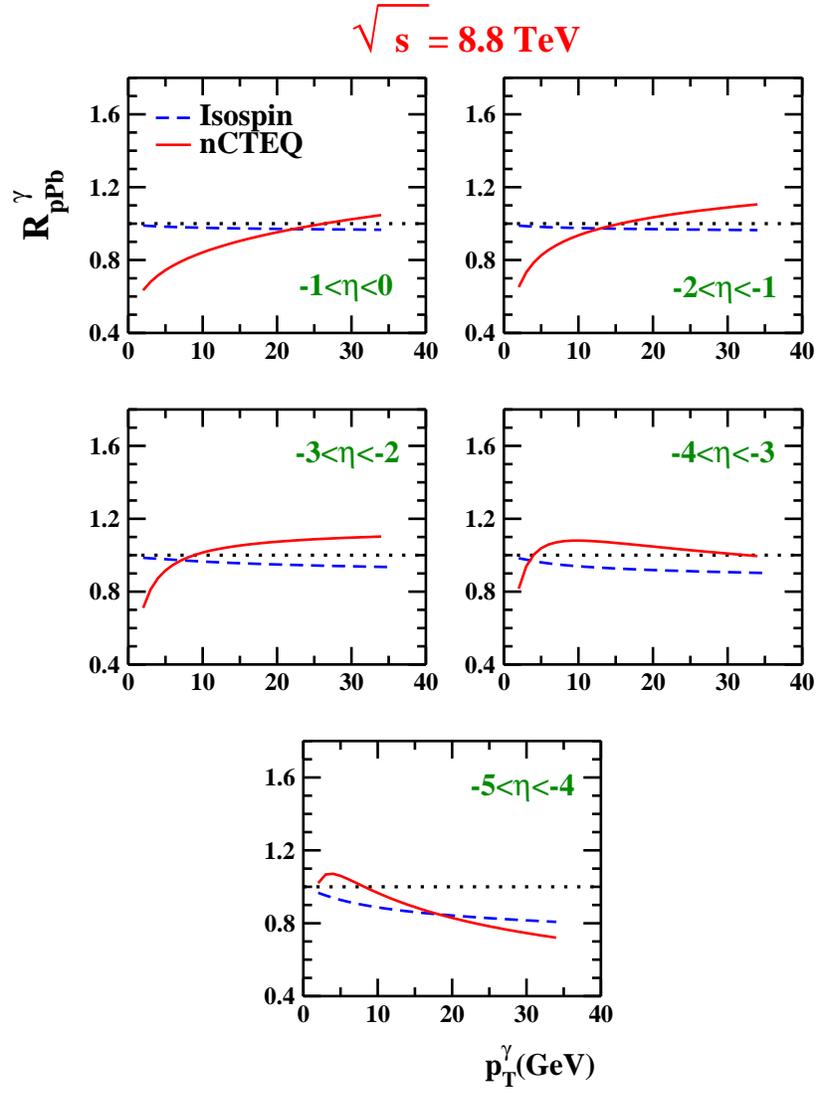}
\caption{Same as Fig.~\ref{fig:fig2}, but for $ \sqrt{s}=8.8 $ TeV.}
\label{fig:fig4}
\end{figure}

The corresponding results for $ \sqrt{s}=8.8 $ TeV have been shown in Fig.~\ref{fig:fig4}. 
This figure clearly indicates that the best kinematic regions for constraining antishadowing nuclear modification at such higher energies are again $ -3 < \eta^\gamma < -2 $ and $ -4 < \eta^\gamma < -3 $, just like the case of $ \sqrt{s}=5.02 $ TeV. However, the difference between the isospin baseline and nCTEQ15 prediction in these regions has decreased at $ \sqrt{s}=8.8 $ TeV in comparison with the
obtained results at $ \sqrt{s}=5.02 $ TeV. Overall, by comparing the three Figs.~\ref{fig:fig2},~\ref{fig:fig3} and~\ref{fig:fig4}, we can conclude that, among center-of-mass energies considered here, the best one for constraining antishadowing nuclear modification in backward regions is $ \sqrt{s}=2.76 $ TeV. Furthermore, at this energy, among different kinematic regions of photon pseudorapidity, the best regions are $ -2 < \eta^\gamma < -1 $ and $ -3 < \eta^\gamma < -2 $.

\subsection{Rapidity normalized distributions}

As mentioned before, the main advantages of quantity $ R_{p\textrm{Pb}}^\gamma $ introduced in 
Eq.~\ref{eq5} are its sensitivity to nuclear modifications of parton densities and also some cancellations of the theoretical and experimental uncertainties. However, looking for similar 
quantities with more benefits is really of interest. It has been indicated that the ratio of distributions for different rapidity regions can be a good option~\cite{Goharipour:2017rjl}. In Refs.~\cite{Gauld:2015yia,Gauld:2016kpd}, the authors showed that using rapidity normalized cross section data from heavy flavor production at LHCb rather than the absolute measurements in a QCD global analysis of PDFs can reduce the scale dependence of the theoretical predictions and then the uncertainties in resulting PDFs. In this section, we are going to examine the possibility of using such quantities to constrain nuclear modifications of PDFs, specially the antishadowing one in our case of study. 

According to what mentioned above, as the first examination we normalize the results obtained for 
$ R_{p\textrm{Pb}}^\gamma $ in Figs.~\ref{fig:fig2},~\ref{fig:fig3} and~\ref{fig:fig4} to the corresponding results of the first rapidity bin $ -1 < \eta^\gamma < 0 $. To be more precise, we define and calculate the rapidity normalized nuclear modification ratio
\begin{equation}
{\cal{R}}_{\eta,i}^\gamma \equiv \frac{R_{p\textrm{Pb}}^\gamma \mid_{\eta_i}}{R_{p\textrm{Pb}}^\gamma \mid_{\eta_1}},
\label{eq6}
\end{equation}
in which, $ \eta_i $ where $ i=2, 3, 4, 5 $, indicate the rapidity bins $ -2 < \eta^\gamma < -1 $,
$ -3 < \eta^\gamma < -2 $, $ -4 < \eta^\gamma < -3 $ and $ -5 < \eta^\gamma < -4 $, respectively, and
$ \eta_1 $ is corresponding to bin $ -1 < \eta^\gamma < 0 $.
 
\begin{figure}[t!]
\centering
\includegraphics[width=0.65\textwidth]{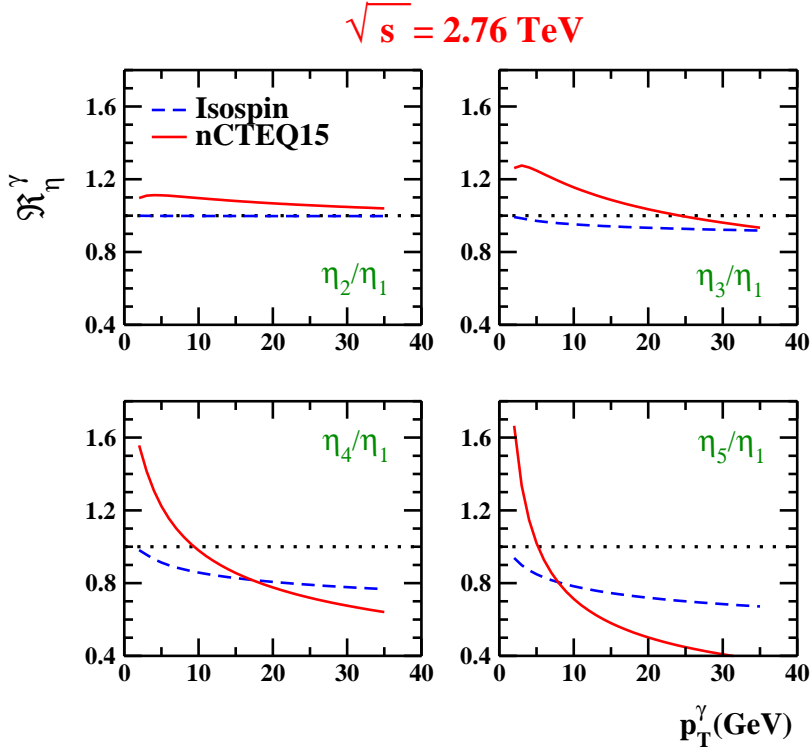}
\caption{The rapidity normalized nuclear modification ratio $ {\cal{R}}_{\eta}^\gamma $ defined in Eq.~\ref{eq6} as a function of $ p_\textrm{T}^\gamma $ at $ \sqrt{s}=2.76 $ TeV and different rapidity regions, with no nuclear modifications in the PDFs (blue dashed curve) and using the nCTEQ15~\cite{Kovarik:2015cma} nuclear modifications (red solid curve).}
\label{fig:fig5}
\end{figure}

\begin{figure}[t!]
\centering
\includegraphics[width=0.65\textwidth]{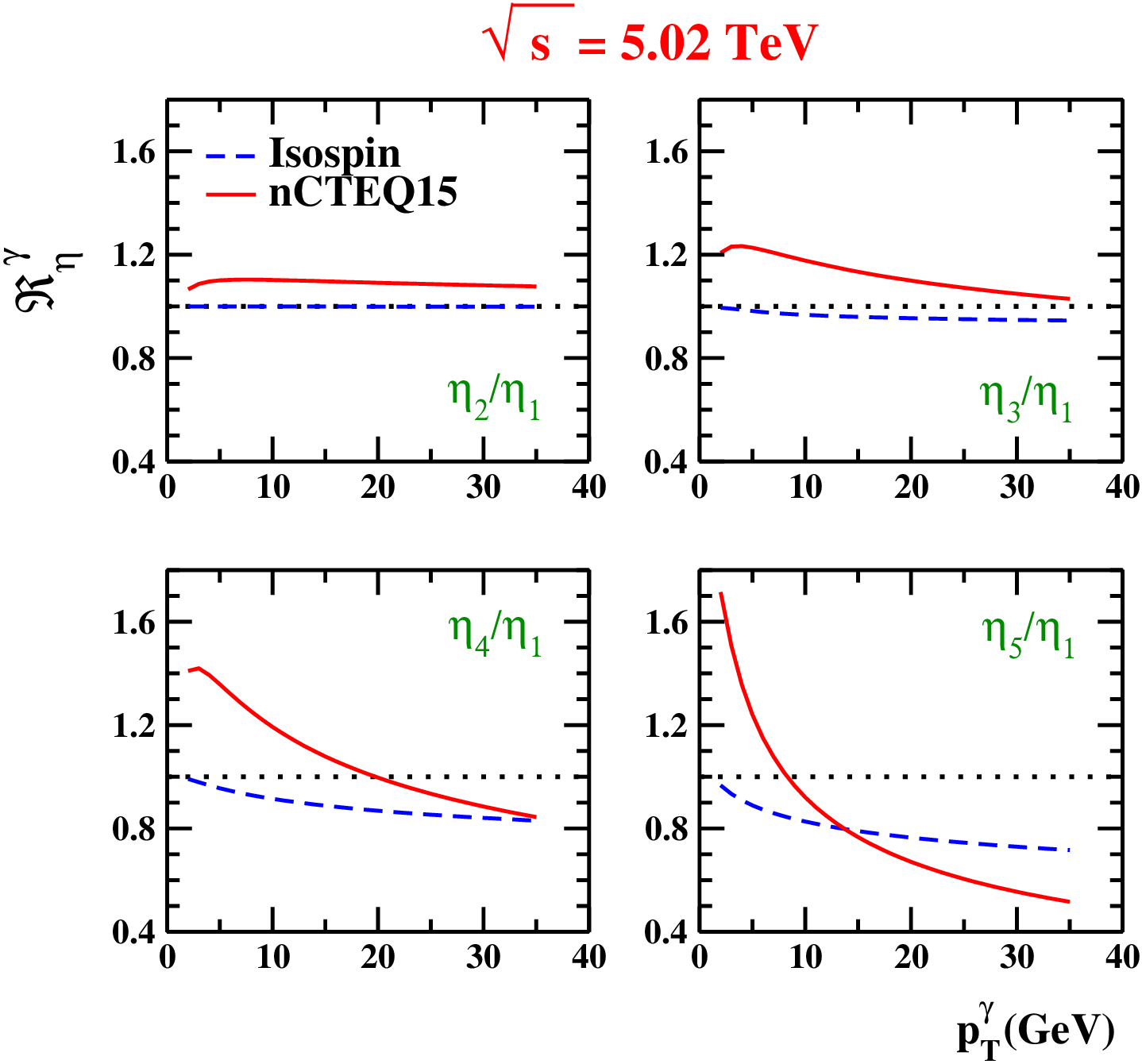}
\caption{Same as Fig.~\ref{fig:fig5}, but for $ \sqrt{s}=5.02 $ TeV.}
\label{fig:fig6}
\end{figure}

The results obtained for different values of center-of-mass energy of $ \sqrt{s}= 2.76, 5.02 $ and 8.8 TeV have been shown in Figs.~\ref{fig:fig5},~\ref{fig:fig6} and~\ref{fig:fig7}, respectively. As can be seen from these figures, there are also some kinematic regions in which the quantity $ {\cal{R}}_{\eta}^\gamma $ defined in Eq.~\ref{eq6} is sensitive to antishadowing nuclear modification. Totally, we can say that the best kinematic regions for constraining antishadowing nuclear
modification using the quantity $ {\cal{R}}_{\eta}^\gamma $ are $ \eta_2/\eta_1 $ for 
$ \sqrt{s}= 2.76 $ TeV, and also $ \eta_3/\eta_1 $ and $ \eta_4/\eta_1 $ for $ \sqrt{s}= 5.02 $ and 8.8 TeV. For the case of $ \sqrt{s}= 8.8 $ TeV, the last panel, in which the nuclear modification ratio $ R_{p\textrm{Pb}}^\gamma $ for $ -5 < \eta^\gamma < -4 $ has been dived to the one for $ -1 < \eta^\gamma < 0 $, can be also a good option if we consider the smaller values of $ p_\textrm{T}^\gamma $. Another point can be noted is that as the value of center-of-mass energy increases, $ {\cal{R}}_{\eta}^\gamma $ shows somewhat more sensitivity to antishadowing nuclear modification. Moreover, note that for the case of $ \sqrt{s}= 2.76 $ TeV, the last panel shows a good sensitivity to the EMC effect nuclear modification. Then, measuring the quantity $ {\cal{R}}_{\eta}^\gamma $ in such kinematic region can also be useful for constraining the EMC effect nuclear modification.

Although the rapidity normalized nuclear modification ratio $ {\cal{R}}_{\eta}^\gamma $ can also be used phenomenologically for investigating the antishadowing nuclear modification according to results obtained, but it is somewhat complicated experimentally compared to usual nuclear modification ratio $ R_{p\textrm{Pb}}^\gamma $. Actually, $ {\cal{R}}_{\eta}^\gamma $ includes two quantities which are expressed in turn as the ratio of cross sections. However, more precisely on this quantity, we find that it can be rewritten as
\begin{equation}
{\cal{R}}_{\eta,i}^\gamma \equiv \frac{R_{p\textrm{Pb}}^\gamma \mid_{\eta_i}}{R_{p\textrm{Pb}}^\gamma \mid_{\eta_1}}=\frac{d\sigma/dp_\textrm{T} \mid_{\eta_i}^{p\textrm{Pb}}}{d\sigma/dp_\textrm{T} \mid_{\eta_1}^{p\textrm{Pb}}}\times \frac{d\sigma/dp_\textrm{T} \mid_{\eta_1}^{pp}}{d\sigma/dp_\textrm{T} \mid_{\eta_i}^{pp}}.
\label{eq7}
\end{equation}
It is observed that the second fraction on the right side of Eq.~\ref{eq7} doesn't include any nuclear modification of PDFs and acts as a factor for the first fraction which includes nuclear modifications both in numerator and denominator. This fact suggests that we can consider the first fraction as a new quantity for exploring the nuclear modification of PDFs. Then, we define the ratio of nuclear cross sections for different rapidity regions as
\begin{equation}
R_{\eta,i}^\gamma \equiv \frac{d\sigma/dp_\textrm{T} \mid_{\eta_i}^{p\textrm{Pb}}}{d\sigma/dp_\textrm{T} \mid_{\eta_1}^{p\textrm{Pb}}},
\label{eq8}
\end{equation}
where the explanation of indexes is as before. Note that since the nuclear cross sections in numerator and denominator of Eq.~\ref{eq8} are corresponding to different rapidity regions, $ R_{\eta}^\gamma $ can be totally sensitive to the nuclear modification of PDFs, though some dependencies may be neutralized. The main advantage of
$ R_{\eta}^\gamma $ is that it dose not require a $ pp $ baseline measurement with the
same $ \sqrt{s} $ that is really an important issue. Actually, in terms of nature, it is like to
the yield asymmetry between the forward and backward rapidities $ Y_{p\textrm{Pb}}^{\textrm{asym}} $ that its usefulness for constraining nuclear PDFs has previously been established~\cite{Helenius:2014qla,Goharipour:2017uic,Aaij:2014pvu}.

The results obtained for $ R_{\eta}^\gamma $ have been shown in Figs.~\ref{fig:fig8},~\ref{fig:fig9} and~\ref{fig:fig10} for different values of center-of-mass energy of $ \sqrt{s}= 2.76, 5.02 $ and 8.8 TeV, respectively. Some interesting conclusions can be achieved from this figures and also by comparing them to the results obtained for $ {\cal{R}}_{\eta}^\gamma $ in Figs.~\ref{fig:fig5},~\ref{fig:fig6} and~\ref{fig:fig7}. First of all, focusing on each result obtained for $ R_{\eta}^\gamma $ at various energies, we find that the sensitivity of this quantity to antishadowing nuclear
modification is decreased as the bins with higher value of absolute rapidity are considered in numerator of $ R_{\eta}^\gamma $. Moreover, as $ p_\textrm{T}^\gamma $ increases, $ R_{\eta}^\gamma $ moves towards the isospin baseline which means that the nuclear modifications in numerator and denominator of $ R_{\eta}^\gamma $ have neutralized each other. By comparing $ R_{\eta}^\gamma $ and
$ {\cal{R}}_{\eta}^\gamma $ results, one can conclude that they have almost similar behavior, but the curves in $ R_{\eta}^\gamma $ have shifted downward. This fact indicates that the $ pp $
fraction on the right side of Eq.~\ref{eq7} really acts as a factor. 
Overall, in this case, the best kinematic regions for constraining antishadowing nuclear
modification are $ \eta_2/\eta_1 $ for $ \sqrt{s}= 2.76 $ TeV, and also $ \eta_3/\eta_1 $ and $ \eta_4/\eta_1 $ for $ \sqrt{s}= 5.02 $ and 8.8 TeV. Another interesting point can be
concluded is that the sensitivity of $ R_{\eta}^\gamma $ to the antishadowing nuclear modification is almost the same size of $ {\cal{R}}_{\eta}^\gamma $. Therefore, it can be confirmed that measuring 
$ R_{\eta}^\gamma $ for investigating the antishadowing area is experimentally preferable rather than $ {\cal{R}}_{\eta}^\gamma $ since there is no need to measure the $ pp $ cross sections with the
same $ \sqrt{s} $.

\subsection{Energy normalized distributions}
Looking for a quantity with more sensitivity to antishadowing nuclear modification
which has the same benefits as before (for example, overall cancellations of the theoretical and experimental uncertainties), in this section we investigate the idea of using the energy normalized distributions. In Ref.~\cite{Gauld:2016kpd}, the authors 
investigated the impact of forward charm production data from the LHCb measurements
at different center-of-mass energies 5, 7, and 13 TeV on the NNPDF3.0 PDFs~\cite{Ball:2014uwa} and demonstrated that including the cross section ratios between data at different energies in the analysis leads to a reduction in uncertainty of gluon PDF. Following this idea, as a first step, we normalize the results obtained for $ R_{p\textrm{Pb}}^\gamma $ in Figs.~\ref{fig:fig3} and~\ref{fig:fig4} at $ \sqrt{s}= $ 5.02 and 8.8 TeV, respectively, to the corresponding results obtained at $ \sqrt{s}=2.76 $ TeV in Fig.~\ref{fig:fig2}
for each rapidity bin, separately. To be more precise, we define and calculate the energy normalized nuclear modification ratio
\begin{equation}
{\cal{R}}_{s,i}^\gamma \equiv \frac{R_{p\textrm{Pb}}^\gamma \mid_{s_i}}{R_{p\textrm{Pb}}^\gamma \mid_{s_1}},
\label{eq9}
\end{equation}
in which, $ s_i $, where $ i= 1, 2 $, indicate different center-of-mass energies 5.02 and 8.8 TeV, respectively, and $ s_1 $ is corresponding to $ \sqrt{s} $= 2.76 TeV.
 
The results obtained for $ {\cal{R}}_{s,1}^\gamma $ and $ {\cal{R}}_{s,2}^\gamma $ have been shown in Figs.~\ref{fig:fig11} and~\ref{fig:fig12}, respectively. As can be seen, in both cases, the quantity 
$ {\cal{R}}_{s}^\gamma $ is sensitive to the shadowing nuclear modification at regions with smaller values of absolute rapidity. However, as the rapidity increases in backward direction, $ {\cal{R}}_{s}^\gamma $ becomes more sensitive to the antishadowing nuclear modification. To be more precise, in both cases, only last two panels can be useful for constraining the antishadowing nuclear modification. Comparing two Figs.~\ref{fig:fig11} and~\ref{fig:fig12}, one can also conclude that with increasing energy in numerator of Eq.~\ref{eq9}, the sensitivity of $ {\cal{R}}_{s}^\gamma $ to the shadowing nuclear modification at smaller values of absolute rapidity, and also its sensitivity to the antishadowing nuclear modification at larger values of absolute rapidity are both intensified.  
A very interesting point is that for $ {\cal{R}}_{s,2}^\gamma $ at rapidity bin
$ -5 < \eta^\gamma < -4 $, the difference between the nCTEQ15 prediction and isospin baseline even reaches over 25\% which is significant compared to previous considered quantities.

Since the energy normalized nuclear modification ratio $ {\cal{R}}_{s}^\gamma $, just like the case of rapidity normalized nuclear modification ratio $ {\cal{R}}_{\eta}^\gamma $ (see Eq.~\ref{eq7}), can be written as a product of two ratios (one containing nuclear cross sections and another without including any nuclear modification), we can remove the $ pp $ baseline and define 
the ratio of nuclear cross sections at different energies as
\begin{equation}
R_{s,i}^\gamma \equiv \frac{d\sigma/dp_\textrm{T} \mid_{s_i}^{p\textrm{Pb}}}{d\sigma/dp_\textrm{T} \mid_{s_1}^{p\textrm{Pb}}},
\label{eq10}
\end{equation}
where the explanation of indexes is as before. However, according to the results obtained for $ {\cal{R}}_{s}^\gamma $, we expect to see some sensitivity to the antishadowing nuclear modification only at higher values of rapidity in backward direction.

The results obtained for $ R_{s,1}^\gamma $ and $ R_{s,2}^\gamma $ have been shown in Figs.~\ref{fig:fig13} and~\ref{fig:fig14}, respectively. Just like the case of $ {\cal{R}}_{s}^\gamma $,
at regions with smaller values of absolute rapidity, the quantities $ R_{s,1}^\gamma $ and $ R_{s,2}^\gamma $ are sensitive to the shadowing nuclear modification and can be used for constraining gluon nPDF which is poorly known at this area. However, as the value of absolute rapidity increases, the spectrum moves towards the isospin baseline and finally some sensitivities to antishadowing nuclear modification are appeared at two rapidity bins
$ -4 < \eta^\gamma < -3 $ and $ -5 < \eta^\gamma < -4 $. Note that unlike $ {\cal{R}}_{s}^\gamma $,
the ratio of nuclear cross sections at different energies $ R_{s}^\gamma $ follows an exponential behavior as the value of absolute rapidity is increased. This fact shows that the ratio of $ pp $
cross sections at different energies (in reverse) balances the overall amount of $ {\cal{R}}_{s}^\gamma $, so that it remains almost around unity (see Figs.~\ref{fig:fig11} and~\ref{fig:fig12}).
Comparing the results obtained for $ R_{s,1}^\gamma $ and $ R_{s,2}^\gamma $, we see that they are similar to a large extent in behavior, though their magnitudes are quite different as expected.
Overall, we can conclude that the measurement of $ R_{s}^\gamma $ in backward direction and at higher values of absolute rapidity and transverse momentum can be very useful for constraining antishadowing nuclear modifications of parton densities especially of the gluon. Moreover, its measurements 
is experimentally preferable rather than $ {\cal{R}}_{s}^\gamma $, since there is no need to measure the $ pp $ cross sections with the same $ \sqrt{s} $.

\subsection{Theoretical uncertainties}

In the previous subsection, we examined various observables to find the best kinematic regions in which the measurement of the isolated prompt photon production in $ p $-Pb collisions at the LHC has most sensitivity to the antishadowing area. In all calculations performed so far, we used only nCTEQ15 nPDFs which have stronger antishadowing nuclear modifications compared with EPS09~\cite{Eskola:2009uj}, DSSZ~\cite{deFlorian:2011fp} and HKN07~\cite{Hirai:2007sx} nPDFs according to Fig.~\ref{fig:fig1}. Note also that nCTEQ15 has greatest uncertainties in comparison with other groups. As a last step, we are going now to present a comparison between various nPDFs and also study the theoretical uncertainties in the results due to the nPDFs uncertainties and scale variations. To this aim, we choose the third panel of Fig.~\ref{fig:fig10} in which the ratio of nuclear cross sections for different rapidity regions ($ \eta_4/\eta_1 $) $ R_{\eta}^\gamma $ defined in Eq.~\ref{eq8} has been plotted as a function of $ p_\textrm{T}^\gamma $ at $ \sqrt{s}=8.8 $ TeV. As mentioned before, such an observable has an advantage to cancel some experimental and theoretical uncertainties and also no need to measure the $ pp $ baseline. Note that the theoretical uncertainties of nCTEQ15 nPDFs in any physical quantity such as the cross section, can be calculated as usual using the 32 error sets of the nCTEQ15 parametrisation (see Ref.~\cite{Goharipour:2017uic}).

The results obtained have been shown in Fig~~\ref{fig:fig15} where the solid, dotted-dashed, dotted-dashed-dashed and dotted-dotted-dashed curves correspond to the nCTEQ15, EPS09, DSSZ and HKN07 predictions, respectively, and the red band corresponds to the nCTEQ15 nPDFs uncertainties. As before, the isospin base line has been shown (dashed curve) so that the antishadowing area can be easily distinguished. As can be seen, although the EPS09, DSSZ, and HKN07 predictions are in good agreement with each other, the nCTEQ15 has significant deviations from them, especially in smaller values of $ p_\textrm{T}^\gamma $. Overall, the EPS09, DSSZ, and HKN07 predictions are placed below the nCTEQ15 prediction in all values of $ p_\textrm{T}^\gamma $, and are not even within the error band of nCTEQ15 at $ p_\textrm{T}^\gamma <20 $ GeV. According to the results obtained, one can simply conclude that the measurements of such quantities at the LHC in appropriate kinematics regions can be really a useful tool in order to accurately determine the antishadowing nuclear modification.

As mentioned before, another important source of the theoretical uncertainties arises from the variation in the renormalization, factorization and fragmentation scales. In all calculations presented so far, we chose $ \mu=M=M_F=p_\textrm{T}^\gamma $. However, it is also of interest to study the changes in the results due to scale variations. Actually, it can demonstrate the effects of higher order calculations and answer to this question that the definition of quantities like the ratio of nuclear cross sections for different rapidity regions $ R_{\eta}^\gamma $, to what extent can resolve the lake of higher order calculations? In this regard, we choose again the third panel of Fig.~\ref{fig:fig10} and recompute $ R_{\eta}^\gamma $ using nCTEQ15 nPDFs, but this time by setting
all scales to be equal and varying them by a factor of 2 around the central value ($ p_\textrm{T}^\gamma $). 

Fig.~\ref{fig:fig16} shows a comparison between the nPDFs and scale uncertainties in $ R_{\eta}^\gamma $ as a function of $ p_\textrm{T}^\gamma $ at $ \sqrt{s}=8.8 $ TeV and for rapidity region $ \eta_4/\eta_1 $. The filled red and hatched green bands represent the nPDFs and scale uncertainties, respectively. Note that the black solid curve corresponds to the results
obtained using the nCTEQ15 central set. The ratios to the nCTEQ15 central prediction have also been shown in the bottom panel. As can be seen, the nPDFs uncertainties are dominant rather than scale uncertainties almost in all ranges of $ p_\textrm{T}^\gamma $, except for smallest values where the scale uncertainties become very large. This figure clearly demonstrates that the definition of quantities like $ R_{\eta}^\gamma $ can significantly reduce the scale uncertainties so that they can be of order of a few percents in some kinematics regions and then negligible compared with the nPDFs uncertainties.

%
\section{Summary and conclusions}\label{sec:five}
The prompt photon production in hadronic collisions has been an interesting subject of research more than three decades, since it provides a powerful tool for testing perturbative QCD and brings
useful information on the gluon PDF and also its nuclear modification. Because of the lake
of experimental data containing nuclei in initial state compared with the case of free nucleon which are suitable for using in a global analysis, our knowledge of nuclear modifications of parton densities is not so well as PDFs. In this work, we performed a comprehensive study of the isolated prompt photon production in $ p $-Pb collisions at backward rapidities to find how its measurements can put constraints on the antishadowing nuclear modification of gluon PDF. In this regard,
we calculated and compared various quantities, including the nuclear modification ratio, the rapidity and energy normalized nuclear modification ratios, and the ratio of nuclear cross sections for different rapidity regions and also different values of center-of-mass energy covered by the LHC to realize which one is most useful. For the case of nuclear modification ratio $ R_{p\textrm{Pb}}^\gamma $, we found that as the absolute value of rapidity increases in backward direction, it moves from the shadowing area to antishadowing and then the EMC effect. We showed that among center-of-mass energies $ \sqrt{s}=2.76, 5.02, 8.8 $ TeV, the best one for constraining antishadowing nuclear modification using $ R_{p\textrm{Pb}}^\gamma $ is the lowest one. Furthermore, at this energy, among different kinematic regions of photon pseudorapidity, the best regions are $ -2 < \eta^\gamma < -1 $ and $ -3 < \eta^\gamma < -2 $.
For the case of rapidity normalized nuclear modification ratio $ {\cal{R}}_{\eta}^\gamma $, we demonstrated that as the value of center-of-mass energy increases, it shows somewhat more sensitivity to antishadowing nuclear modification, so that the best kinematic regions for constraining antishadowing are $ \eta_3/\eta_1 $ and $ \eta_4/\eta_1 $ for $ \sqrt{s}= 8.8 $ TeV.
Meanwhile, we found that measuring $ {\cal{R}}_{\eta}^\gamma $ at lower center-of-mass energy $ \sqrt{s}=2.76 $ TeV and rapidity region $ \eta_5/\eta_1 $ can be also used for constraining the EMC effect nuclear modification. Then, we calculated the ratio of nuclear cross sections for different rapidity regions $ R_{\eta}^\gamma $ at various energies. We found that the sensitivity of this quantity to antishadowing nuclear modification is decreased as the bins with higher value of absolute rapidity are considered. We indicated that $ R_{\eta}^\gamma $ and $ {\cal{R}}_{\eta}^\gamma $ have almost similar behavior, but the curves in $ R_{\eta}^\gamma $ are shifted downward. However, the sensitivity of $ R_{\eta}^\gamma $ to the antishadowing nuclear modification is almost the same size of $ {\cal{R}}_{\eta}^\gamma $. Therefore, we concluded that measuring $ R_{\eta}^\gamma $ is experimentally preferable rather than $ {\cal{R}}_{\eta}^\gamma $ since there is no need to measure the $ pp $ cross sections with the same $ \sqrt{s} $. For the case of energy normalized nuclear modification ratio
$ {\cal{R}}_{s}^\gamma $, we show that it is sensitive to the shadowing nuclear modification at regions with smaller values of absolute rapidity. However, as the rapidity increases in backward direction, $ {\cal{R}}_{s}^\gamma $ becomes more sensitive to the antishadowing nuclear modification,
so that the best kinematic regions are $ -4 < \eta^\gamma < -3 $ and $ -5 < \eta^\gamma < -4 $.
Moreover, we found that with increasing energy in numerator of $ {\cal{R}}_{s}^\gamma $, its sensitivity to the shadowing and antishadowing nuclear modifications at smaller and larger values of absolute rapidity, respectively, is intensified in both cases. Finally, we calculated the ratio of nuclear cross sections at different energies $ R_{s}^\gamma $ and found that it behaves to a large extent as $ {\cal{R}}_{s}^\gamma $, though its magnitude is quite different. To be more precise, it can be used for constraining the shadowing and antishadowing nuclear modifications at smaller and larger values of absolute rapidity, respectively. However, we confirm that its measurement in backward direction and at higher values of absolute rapidity and transverse momentum
is experimentally preferable rather than $ {\cal{R}}_{s}^\gamma $ for constraining the antishadowing area, since there is no need to measure the $ pp $ cross sections with the same $ \sqrt{s} $. We also presented a comparison between various nPDFs and also studied the theoretical uncertainties due to the nPDFs uncertainties and scale variations. We showed that, for $ R_{\eta}^\gamma $ as an example, the EPS09, DSSZ, and HKN07 predictions are in good agreement with each other, but the nCTEQ15 has significant deviations from them, especially in smaller values of $ p_\textrm{T}^\gamma $. We concluded that the measurements of such quantities at the LHC in appropriate kinematics regions can be really a useful tool in order to accurately determine the antishadowing nuclear modification. Moreover, we demonstrated that the nPDFs uncertainties are dominant rather than scale uncertainties almost in all ranges of $ p_\textrm{T}^\gamma $. We confirm that the definition of quantities like $ R_{\eta}^\gamma $ can significantly reduce the scale uncertainties, so that they can be of order of a few percents in some kinematics regions and then negligible compared with the nPDFs uncertainties.

%
\acknowledgments
Authors thank Hamzeh Khanpour for useful discussions and comments.
Muhammad Goharipour also thanks the School of Particles and Accelerators,
Institute for Research in Fundamental Sciences (IPM)
for financial support provided for this research.
%

%

\newpage

\begin{figure}[t!]
\centering
\includegraphics[width=0.65\textwidth]{fig7}
\caption{Same as Fig.~\ref{fig:fig5}, but for $ \sqrt{s}=8.8 $ TeV.}
\label{fig:fig7}
\end{figure}
\begin{figure}[t!]
\centering
\includegraphics[width=0.65\textwidth]{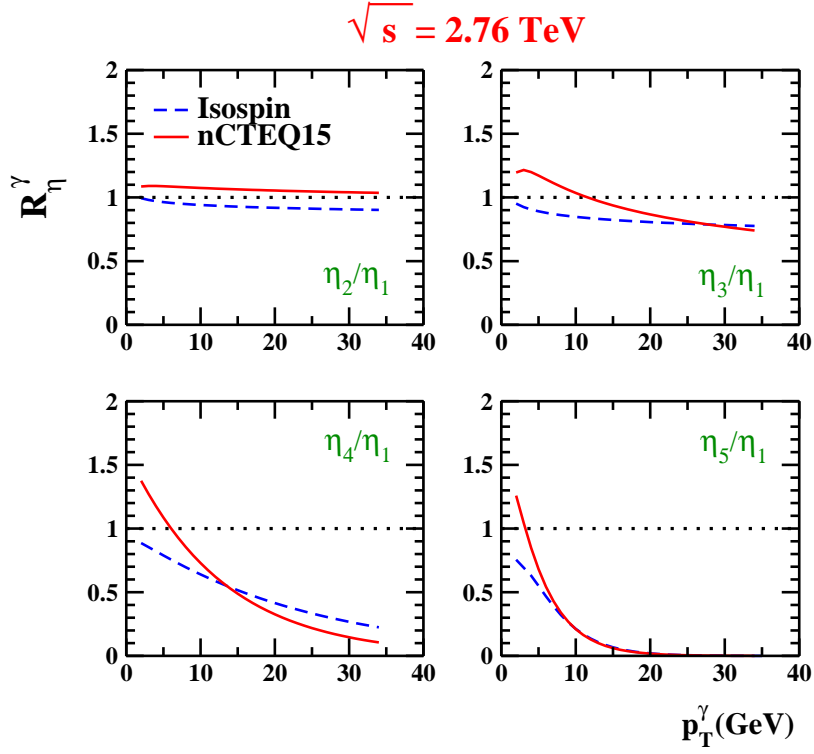}
\caption{The ratio of nuclear cross sections for different rapidity regions $ R_{\eta}^\gamma $ defined in Eq.~\ref{eq8} as a function of $ p_\textrm{T}^\gamma $ at $ \sqrt{s}=2.76 $ TeV, with no nuclear modifications in the PDFs (blue dashed curve) and using the nCTEQ15~\cite{Kovarik:2015cma} nuclear modifications (red solid curve).}
\label{fig:fig8}
\end{figure}
\begin{figure}[t!]
\centering
\includegraphics[width=0.65\textwidth]{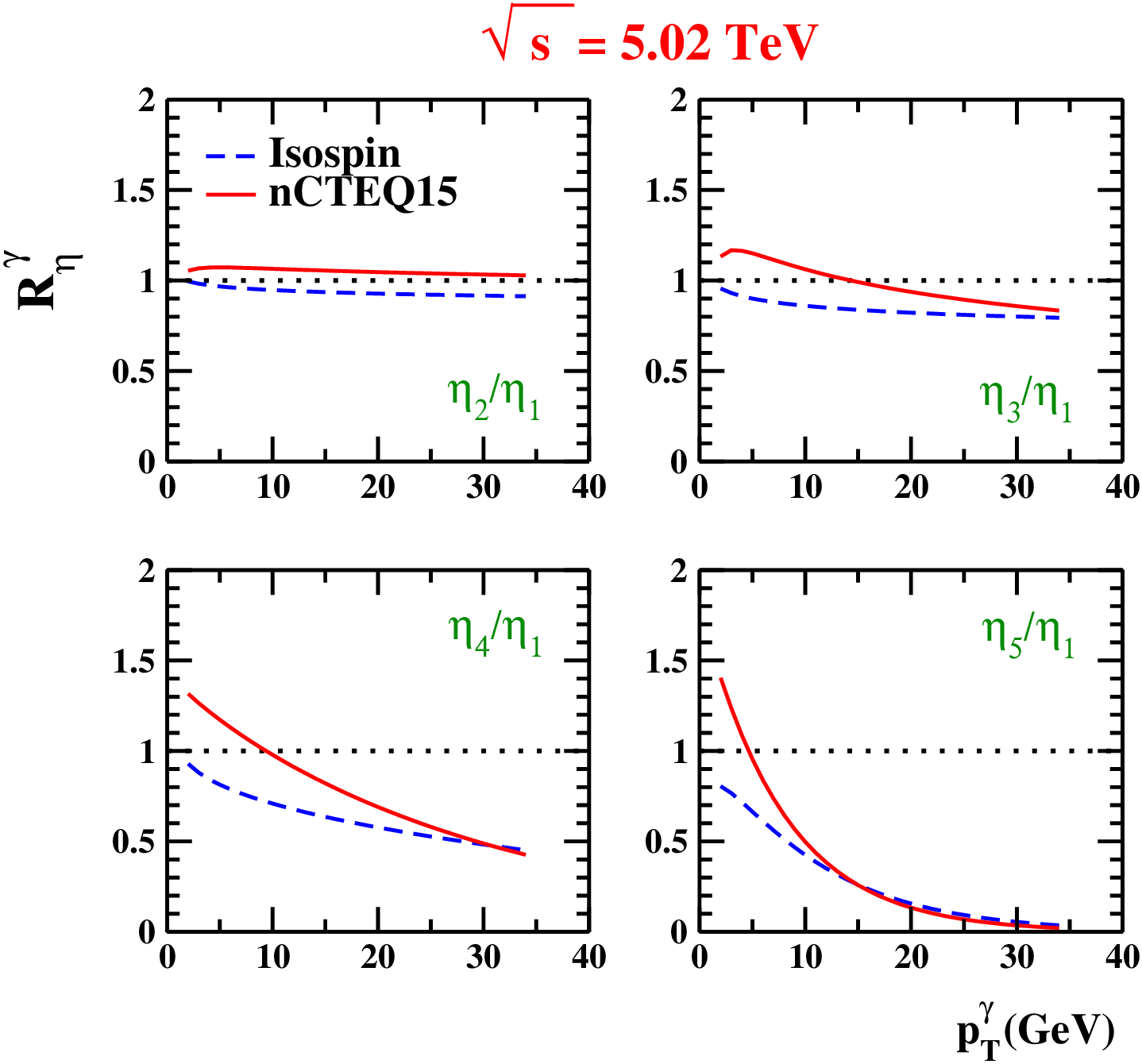}
\caption{Same as Fig.~\ref{fig:fig8}, but for $ \sqrt{s}=5.02 $ TeV.}
\label{fig:fig9}
\end{figure}
\begin{figure}[t!]
\centering
\includegraphics[width=0.65\textwidth]{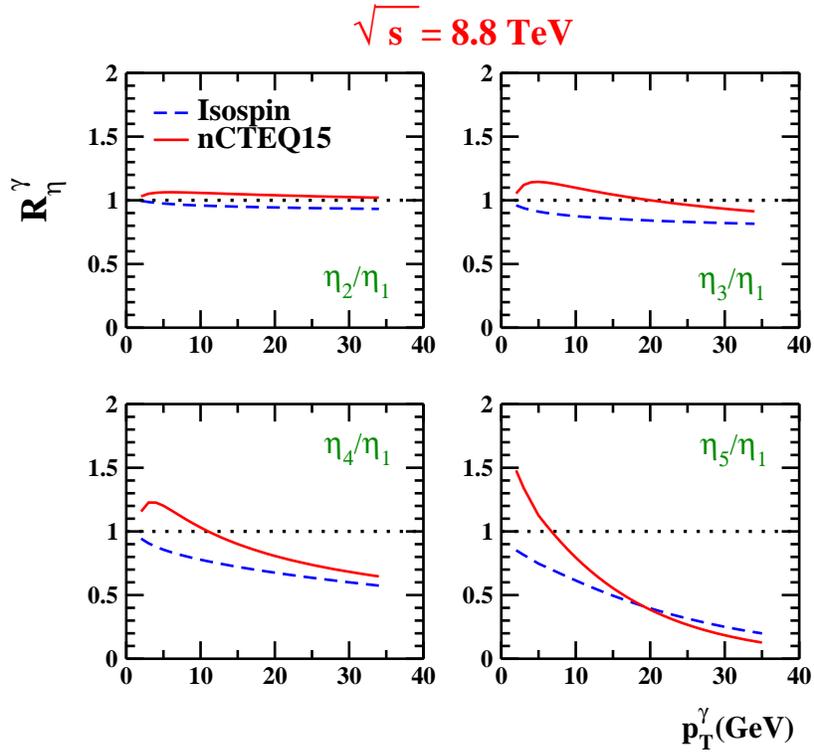}
\caption{Same as Fig.~\ref{fig:fig8}, but for $ \sqrt{s}=8.8 $ TeV.}
\label{fig:fig10}
\end{figure}
\begin{figure}[t!]
\centering
\includegraphics[width=0.65\textwidth]{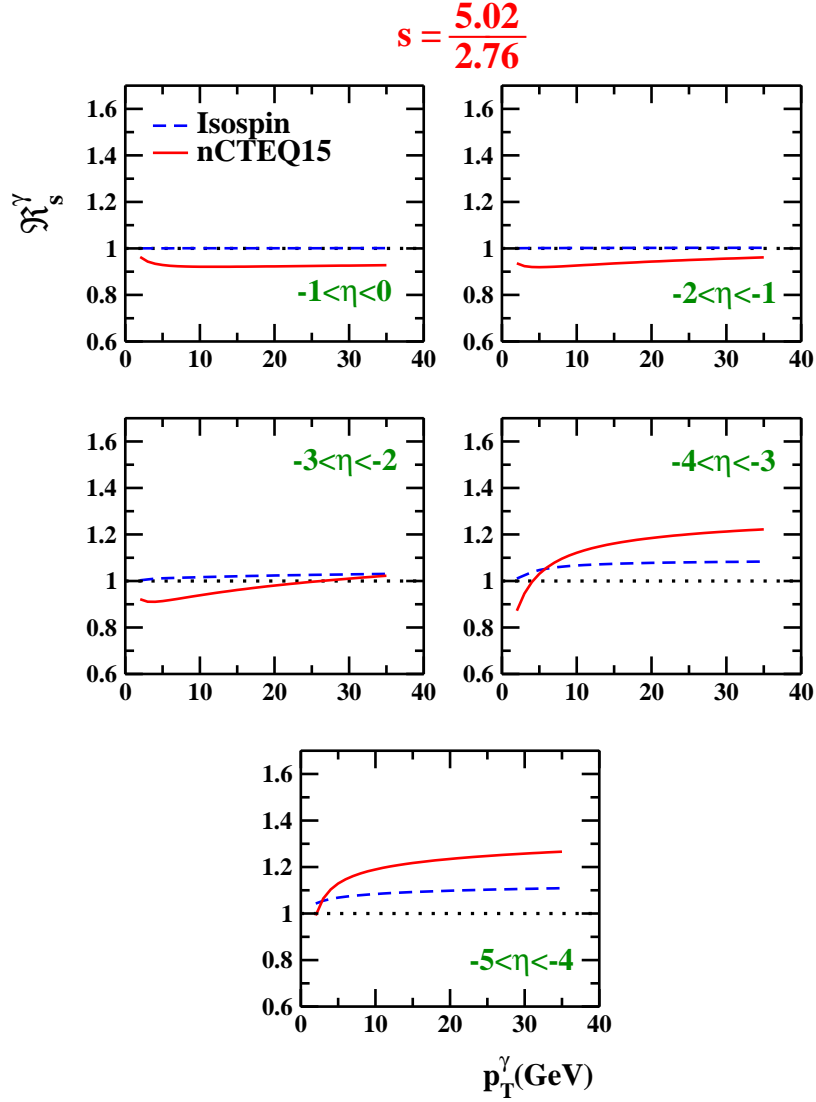}
\caption{The energy normalized nuclear modification ratio $ {\cal{R}}_{s,1}^\gamma $ defined in Eq.~\ref{eq9} as a function of $ p_\textrm{T}^\gamma $ at different rapidity regions, with no nuclear modifications in the PDFs (blue dashed curve) and using the nCTEQ15~\cite{Kovarik:2015cma} nuclear modifications (red solid curve).}
\label{fig:fig11}
\end{figure}
\begin{figure}[t!]
\centering
\includegraphics[width=0.65\textwidth]{fig12}
\caption{Same as Fig.~\ref{fig:fig11}, but for $ {\cal{R}}_{s,2}^\gamma $.}
\label{fig:fig12}
\end{figure}
\begin{figure}[t!]
\centering
\includegraphics[width=0.65\textwidth]{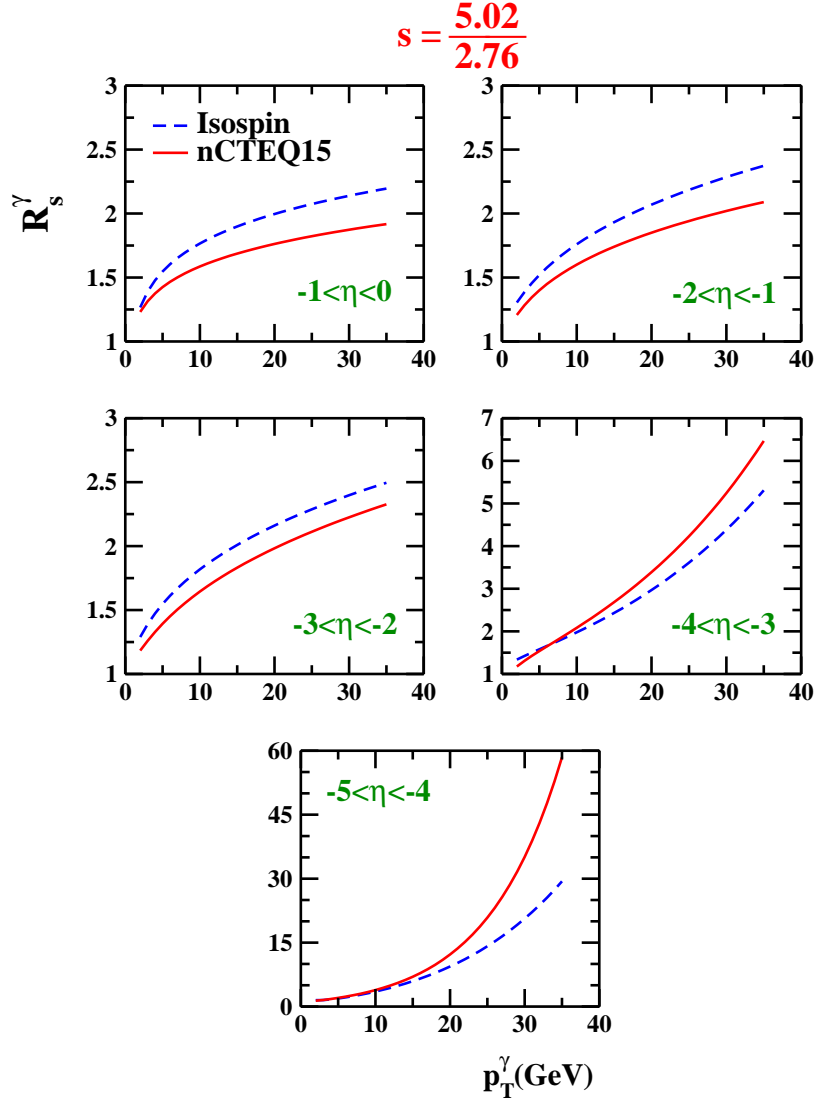}
\caption{The ratio of nuclear cross sections at different energies $ R_{s,1}^\gamma $ defined in Eq.~\ref{eq10} as a function of $ p_\textrm{T}^\gamma $ at at different rapidity regions, with no nuclear modifications in the PDFs (blue dashed curve) and using the nCTEQ15~\cite{Kovarik:2015cma} nuclear modifications (red solid curve).}
\label{fig:fig13}
\end{figure}
\begin{figure}[t!]
\centering
\includegraphics[width=0.65\textwidth]{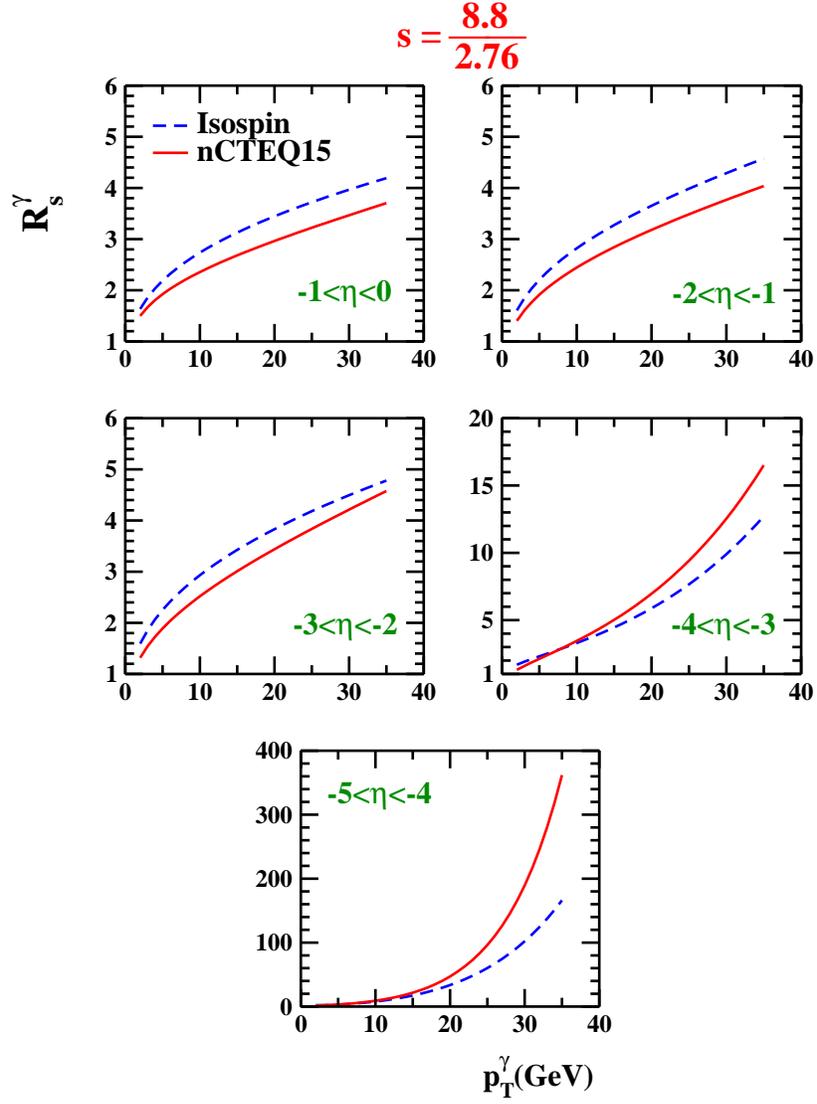}
\caption{Same as Fig.~\ref{fig:fig13}, but for $ R_{s,2}^\gamma $.}
\label{fig:fig14}
\end{figure}
\begin{figure}[t!]
\centering
\includegraphics[width=0.65\textwidth]{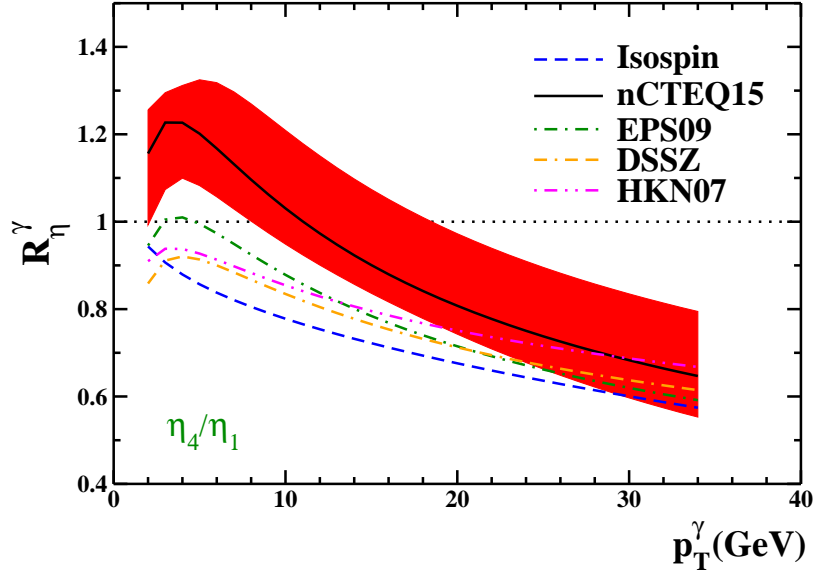}
\caption{The ratio of nuclear cross sections for different rapidity regions ($ \eta_4/\eta_1 $) $ R_{\eta}^\gamma $ defined in Eq.~\ref{eq8} as a function of $ p_\textrm{T}^\gamma $ at $ \sqrt{s}=8.8 $ TeV, with no nuclear modifications in the PDFs (dashed) and using the nCTEQ15 (solid), EPS09 (dotted-dashed), DSSZ (dotted-dashed-dashed) and HKN07 (dotted-dotted-dashed) nuclear modifications. The red band corresponds to the nCTEQ15 nPDF uncertainties.}
\label{fig:fig15}
\end{figure}
\begin{figure}[t!]
\centering
\includegraphics[width=0.65\textwidth]{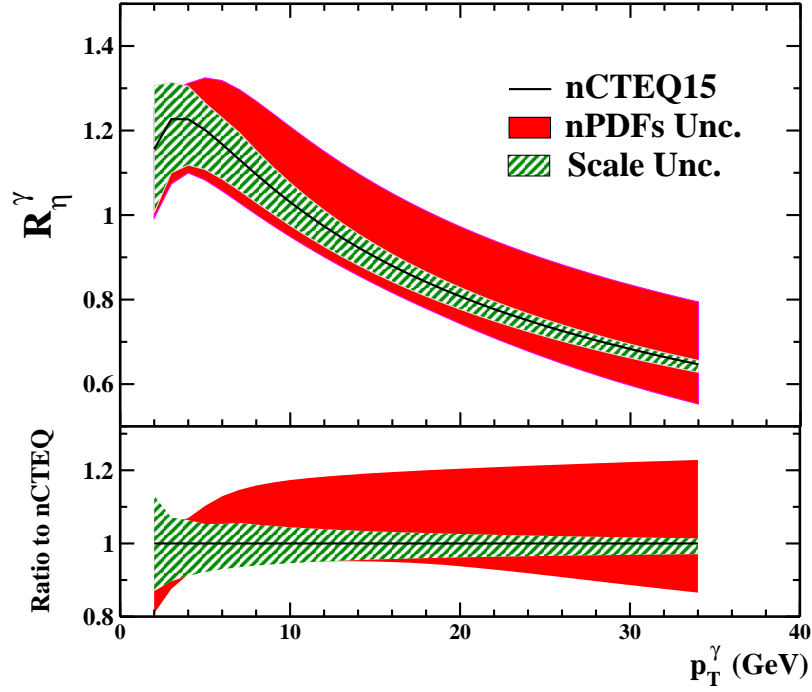}
\caption{A comparison between the nPDFs and scale uncertainties in $ R_{\eta}^\gamma $ (see Eq.~\ref{eq8}) as a function of $ p_\textrm{T}^\gamma $ at $ \sqrt{s}=8.8 $ TeV and for rapidity region $ \eta_4/\eta_1 $. The black solid curve corresponds to the results
obtained using the nCTEQ15 central set. The filled red and hatched green bands represent the nPDFs and scale uncertainties, respectively. The bottom panel shows the
ratios to the nCTEQ15 central prediction.}
\label{fig:fig16}
\end{figure}

\end{document}